\documentclass[prl,twocolumn,superscriptaddress,showpacs]{revtex4}
\usepackage{graphicx,amsmath,amssymb,bm}

\newcommand{\be}{\begin{equation}}
\newcommand{\ee}{\end{equation}}
\newcommand{\kf}{k_{\rm F}}
\newcommand{\fmiq}{\, \text{fm}^{-3}}
\newcommand{\mev}{\, \text{MeV}}
\newcommand{\gevi}{\, \text{GeV}^{-1}}
\newcommand{\znbbeq}{0\nu\beta\beta}
\newcommand{\znbb}{$\znbbeq$}
\newcommand{\tnbb}{$2\nu\beta\beta$}

\begin{document}

\title{Chiral two-body currents in nuclei: \\
Gamow-Teller transitions and neutrinoless double-beta decay}

\author{J.\ Men\'{e}ndez}
\affiliation{Institut f\"ur Kernphysik, 
Technische Universit\"at Darmstadt, 
64289 Darmstadt, Germany}
\affiliation{ExtreMe Matter Institute EMMI, 
GSI Helmholtzzentrum f\"ur Schwerionenforschung GmbH, 
64291 Darmstadt, Germany}
\author{D.\ Gazit}
\affiliation{Racah Institute of Physics, 
The Hebrew University, 
91904 Jerusalem, Israel}
\author{A.\ Schwenk}
\affiliation{ExtreMe Matter Institute EMMI, 
GSI Helmholtzzentrum f\"ur Schwerionenforschung GmbH, 
64291 Darmstadt, Germany}
\affiliation{Institut f\"ur Kernphysik, 
Technische Universit\"at Darmstadt, 
64289 Darmstadt, Germany}

\begin{abstract}
We show that chiral effective field theory (EFT) two-body currents
provide important contributions to the quenching of
low-momentum-transfer Gamow-Teller transitions, and use chiral EFT
to predict the momentum-transfer dependence that is probed in
neutrino-less double-beta (\znbb) decay. We then calculate for the
first time the \znbb\ decay operator based on chiral EFT currents
and study the nuclear matrix elements at successive orders. The
contributions from chiral two-body currents are significant and
should be included in all calculations.
\end{abstract}


\maketitle

Weak interaction processes provide unique probes of the physics of
nuclei and fundamental symmetries, and play a central role in
astrophysics~\cite{weak}. The structure of strongly interacting
systems is explored with $\beta$ decays and weak transitions.
Superallowed decays allow high precision tests of the Standard Model,
and \znbb\ decays probe the nature of neutrinos, their hierarchy and
mass. Weak processes mediate nuclear reactions that drive stellar
evolution, supernovae and nucleosynthesis.

Surprisingly, key aspects of well-known decays remain a puzzle. In
particular, when calculations of Gamow-Teller (GT) transitions of the
spin--isospin-lowering operator $g_A {\bm \sigma} \tau^{-}$ are
confronted with experiment, some degree of renormalization, or
``quenching'' $q$, of the axial coupling $g_A^{\rm eff} = q g_A$ is
needed. Compared to the single-nucleon value 
$g_A=1.2695(29)$, the GT term seems to be weaker in nuclei. This was
first conjectured in studies of $\beta$-decay rates, with a typical
$q \approx 0.75$ in shell-model (SM) calculations~\cite{WildenthalMP} and
other many-body approaches~\cite{QRPA_EDF}. In view of the significant
effect on weak reaction rates, it is no surprise that this suppression
has been the target of many theoretical works. It is also a
major uncertainty for \znbb\ decay nuclear matrix elements (NMEs),
which probe larger momentum transfers of order the pion mass, $p
\sim m_\pi$, where the renormalization could be different. Here we
revisit this puzzle based on chiral EFT currents.

\begin{figure}
\begin{center}
\includegraphics[scale=0.5,clip=]{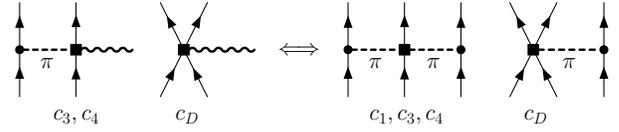}
\end{center}
\vspace*{-4mm}
\caption{Chiral 2b currents and 3N force contributions.}
\label{currents}
\vspace*{-3mm}
\end{figure}

\begin{figure}
\begin{center}
\includegraphics[scale=0.34,clip=]{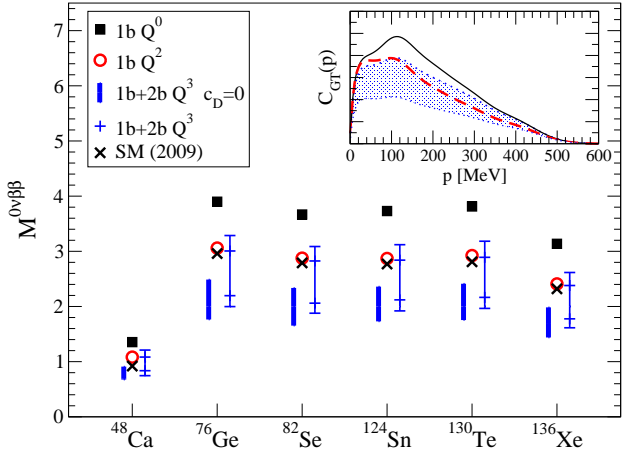}
\end{center}
\vspace*{-5mm}
\caption{(Color online) Nuclear matrix elements $M^{\znbbeq}$ for
\znbb\ decay. At order $Q^0$, the NMEs include only the leading
$p=0$ axial and vector 1b currents. At the next order, all $Q^2$
1b-current contributions not suppressed by parity are taken into
account. At order $Q^3$, the thick bars are predicted from the
long-range parts of 2b currents ($c_D=0$). The thin bars estimate
the theoretical uncertainty from the short-range coupling $c_D$
by taking an extreme range for the quenching (see text). For 
comparison, we show the SM results of Ref.~\cite{Menendez} based 
on phenomenological 1b currents. The inset (representative for 
$^{136}$Xe) shows that the GT part, $M^{\znbbeq}_{\rm GT} = \int 
dp \, C_{\rm GT}(p)$, is dominated by $p \sim 100 \mev$.\label{nme}}
\vspace*{-5mm}
\end{figure}

\begin{figure}
\begin{center}
\includegraphics[scale=0.335,clip=]{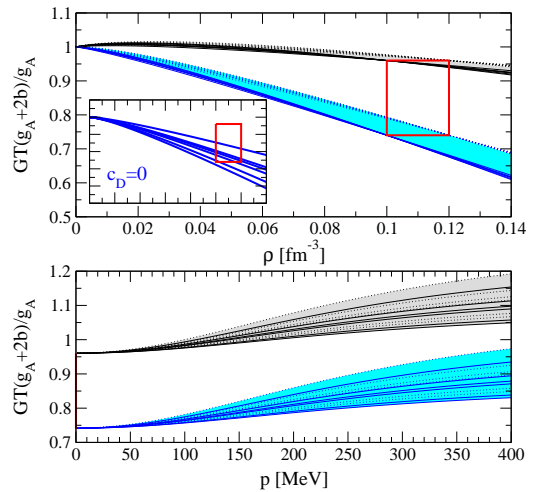}
\end{center}
\vspace*{-5mm}
\caption{(Color online) Top panel: $g_A$ plus 2b-current contributions
for $p=0$ GT transitions normalized to $g_A$ as a function of
density $\rho$. The boundaries of the box are given by $q=0.74/0.96$
and $\rho=0.10...0.12 \fmiq$. The different curves correspond to the
couplings in Table~\ref{table}, with shaded regions for the density
range. Intermediate $q$ values would lie between these regions.
The inset shows the quenching predicted by the
long-range parts of 2b currents only ($c_D=0$). Bottom panel: Same
as top, but as a function of momentum transfer $p$ for
empirical/smaller quenching $q=0.74/0.96$.\label{GT_1b2b}}
\vspace*{-4mm}
\end{figure}

Chiral EFT provides a systematic basis for nuclear forces and
consistent electroweak currents~\cite{RMP,Park}, where pion couplings
contribute both to the electroweak axial current
and to nuclear interactions. This is already seen at leading
order: $g_A$ determines the axial one-body (1b) current and the
one-pion-exchange potential. Two-body (2b) currents, also known as
meson-exchange currents, enter at higher order, just like
three-nucleon (3N) forces~\cite{RMP}. As shown in Fig.~\ref{currents},
the leading axial contributions are due to long-range one-pion-exchange and
short-range parts~\cite{Park}, with couplings $c_3, c_4$ and $c_D$,
which also enter the leading 3N (and subleading NN) 
forces~\cite{RMP,Gazit}. Although the importance of 2b currents
is known from phenomenological studies~\cite{MEC}, chiral currents and
the consistency with nuclear forces have only been explored in the
lightest nuclei~\cite{Park,Gazit,light}. In this Letter, we present
first calculations for GT transitions and for the \znbb\ decay
operator based on chiral EFT currents. A preview of the NMEs
(Fig.~\ref{nme}) and the quenching of $g_A$ (Fig.~\ref{GT_1b2b}) shows
the great importance of using chiral 2b currents in nuclei.

At low energies, the coupling to weak probes is given by
the current-current interaction, $H_W = \frac{G_F}{\sqrt{2}} \int d^3{\bf r}
\, e^{-i {\bf p} \cdot {\bf r}} j_{L \mu} J_L^{\mu \dagger} + {\rm
h.c.}$, where $G_F$ is the Fermi constant, ${\bf p}$ the momentum
transferred from nucleons to leptons, and $j_{L \mu}$ the leptonic
current of an electron coupled to a left-handed electron
neutrino~\cite{weak}. In chiral EFT, the nuclear current $J_L^{\mu
\dagger}$ is organized in an expansion in powers of momentum $Q\sim
m_\pi$ over a breakdown scale $\Lambda_{\rm b} \sim 500
\mev$. Consistently with nuclear forces~\cite{RMP}, we count the
nucleon mass as a large scale, corresponding numerically to $Q/m \sim
(Q/\Lambda_{\rm b})^2$, so that the leading relativistic $1/m$
corrections are of order $Q^2$, and $1/m^2$ terms of order $Q^4$.  To
order $Q^2$ (and also $Q^3$ in this counting), the 1b current,
$J_L^{\mu\dagger}({\bf r}) = \sum_{i=1}^{A} \tau_{i}^{-} \bigl[
\delta^{\mu0} J_{i,{\rm 1b}}^{0} - \delta^{\mu k}J_{i,{\rm 1b}}^{k}
\bigr] \delta({\bf r}-{\bf r}_i)$,
has temporal and spatial parts in momentum space~\cite{Park}:
\begin{align}
J_{i,{\rm 1b}}^{0}(p^{2}) &=  g_{V}(p^{2}) - g_{A} \, \frac{{\bf P}
\cdot {\bm \sigma}_{i}}{2 m} + g_{P}(p^{2}) \, \frac{E \, ({\bf p}
\cdot {\bm \sigma}_{i})}{2 m} \,, 
\label{Jt} \\
{\bf J}_{i,{\rm 1b}}(p^{2}) &= g_{A}(p^{2}) \, {\bm \sigma}_{i}
- g_{P}(p^{2}) \, \frac{{\bf p} \, ({\bf p} \cdot {\bm \sigma}_{i})}{2m}
\nonumber \\
& + i (g_{M}+g_{V}) \, \frac{{\bm \sigma}_{i}\times {\bf p}}{2 m}
- g_{V} \, \frac{{\bf P}}{2 m} \,,
\label{Js}
\end{align}
where $E=E_i-E_i', {\bf p}={\bf p}_i-{\bf p}_i'$, and ${\bf P}={\bf
p}_i+{\bf p}_i'$; and vector ($V$), axial ($A$), pseudo-scalar
($P$), and magnetic ($M$) couplings, $g_{V}(p^{2})$, $g_{A}(p^{2})$,
$g_{P}(p^{2})$, and $g_{M}(p^{2})$~\cite{dipole}. In
chiral EFT, the $p$ dependence is due to loop corrections and
pion propagators, to order~$Q^2$:
$g_{V,A}(p^{2}) = g_{V,A} \, (1-2 \frac{p^2}{\Lambda_{V,A}^2})$, with
$g_V=1$, $\Lambda_V=850 \mev$, 
$\Lambda_A = 2 \sqrt{3}/r_A = 1040 \mev$;
$g_P(p^2) = \frac{2 g_{\pi p n} F_\pi}{m_\pi^2+{\bf p}^2} - 4 \, g_A(p^2)
\frac{m}{\Lambda_A^2}$ and $g_M=\mu_p-\mu_n=3.70$, with pion decay 
constant $F_\pi = 92.4 \mev$, 
$m_{\pi}=138.04 \mev$, and $g_{\pi p n} = 13.05$~\cite{Bernard}.

At leading order $Q^0$, only the momentum-independent $g_A$ and $g_V$
terms contribute. They give rise to 
$p \lesssim 1 \mev$ GT and Fermi ($\tau^{-}$) single-$\beta$ and
\tnbb\ decay. On the other hand, when studying processes that probe
larger momentum transfers, such as \znbb\ decay with $p \sim 100
\mev$, terms of order $Q^2$ need to be included. For \znbb\ decay, the
$Q^0$ terms are still most important and the axial term dominates.
In SM calculations, one has
$M^{0\nu\beta\beta}_{Q^0,{\rm axial}}/M^{0\nu\beta\beta} \approx 1.20$ and
$M^{0\nu\beta\beta}_{Q^0,{\rm vector}}/M^{0\nu\beta\beta} \approx
0.15$ compared to the final $M^{0\nu\beta\beta}$~\cite{Menendez}. Other 
calculations yield similar ratios, also for the $Q^2$
terms, except the vector term can be significantly
larger~\cite{Simkovic,Kortelainen}.

Among the $Q^2$ terms, form-factor-type (ff) contributions and the
$g_P$ part of ${\bf J}_{i,{\rm 1b}}$ dominate. The pseudoscalar term
is important, because $p \,g_P(p^2) \approx 7.9$ for $p \sim 100 \mev$
in \znbb\
decay. They reduce the NMEs: $M^{0\nu\beta\beta}_{\rm ff
}/M^{0\nu\beta\beta} \approx -0.20$ and
$M^{0\nu\beta\beta}_{g_P}/M^{0\nu\beta\beta} \approx
-0.20$~\cite{Menendez}. The remaining $Q^2$ terms are odd under
parity, so they require either a $P$-wave electron (whose phase space
is suppressed~\cite{Tomoda} by 
$\approx 0.03-0.06$ for
\znbb\ decay candidates)
or another
odd-parity term to connect $0^+$ states. Therefore, the ${\bf P}$ and
$E$ terms in Eqs.~(\ref{Jt}) and (\ref{Js}) can be neglected, and 
only the term with the large $g_{M}+g_{V} = 4.70$ is
kept, leading to a small $\approx 5\%$ contribution~\cite{Menendez}.

At order $Q^3$, 2b currents enter in chiral EFT~\cite{Park}. These
include vector spatial, axial temporal, and axial spatial
parts~\cite{vectortemp}. The first two are odd under parity, and
therefore can be neglected. Consequently, for the cases studied here,
the dominant weak 2b currents only have an axial spatial component,
${\bf J}^{\rm axial}_{\rm 2b} = \sum^A_{i<j} {\bf J}_{ij}$,
with~\cite{Park}
\begin{align}
&{\bf J}_{12} = -\frac{g_A}{F^2_{\pi}} \Bigl[ 2 d_1 ({\bm \sigma}_1
\tau_1^- + {\bm \sigma}_2 \tau_2^-) + d_2 \, {\bm \sigma}_{\times} 
\tau^-_{\times} \Bigr] \nonumber \\[1mm]
-&\,\frac{g_A}{2 F^2_{\pi}} \frac{1}{m^2_{\pi}+{\bf k}^2}
\Bigl[ \Bigl(c_4+\frac{1}{4 m} \Bigr) \, {\bf k} \times
({\bm \sigma}_{\times} \times {\bf k}) \, \tau^-_{\times} \nonumber \\[1mm]
+&\,4 c_3 {\bf k} \cdot ({\bm \sigma}_1 \tau_1^- + {\bm \sigma}_2
\tau_2^-) {\bf k} - \frac{i}{2m} {\bf k} \cdot ({\bm \sigma}_1-{\bm
\sigma}_2) {\bf q} \, \tau^-_{\times} \Bigr] ,
\label{2b}
\end{align}
where $\tau_{\times}^-=(\tau_1\times\tau_2)^-$ and the same for ${\bm
\sigma}_{\times}$, ${\bf k}=\frac{1}{2}({\bf p}'_2-{\bf p}_2-{\bf
p}'_1+{\bf p}_1)$ and ${\bf q}=\frac{1}{4}({\bf p}_1+{\bf p}'_1-{\bf
p}_2-{\bf p}'_2)$. Equation~(\ref{2b}) includes contributions
from the one-pion-exchange $c_3, c_4$ parts
and from the short-range couplings $d_1, d_2$, where due to the Pauli
principle only the combination $d_1+2 d_2 = c_D/(g_{A}\Lambda_{\chi})$ 
enters [with $\Lambda_{\chi} = 700 \mev$].

We study the impact of chiral 2b currents in nuclei at the
normal-ordered 1b level by summing the second nucleon over occupied
states in a spin and isospin symmetric reference state or core: ${\bf
J}^{\rm eff}_{i,{\rm 2b}} = \sum_j (1-P_{ij}) {\bf J}_{ij}$, where
$P_{ij}$ is the exchange operator.
The normal-ordered 1b level is expected to be
a very good approximation in medium-mass and heavy nuclei, because of
phase space arguments for normal Fermi systems at low
energies~\cite{Fermi}. 
This approximation has also been explored for chiral 2b currents
in nuclear matter~\cite{Parknm}, but 
limited to long wavelengths and without connecting 2b currents and
nuclear forces. Taking a Fermi gas approximation for the core
and neglecting tensor-like terms $({\bf k} \cdot {\bm
\sigma} \, {\bf k} - \frac{1}{3} k^2 {\bm \sigma}) \tau^-$, we obtain
the normal-ordered 1b current:
\begin{align}
{\bf J}^{\rm eff}_{i,{\rm 2b}} &= 
- g_A {\bm \sigma}_i \tau_i^- \frac{\rho}{F^2_\pi} 
\biggl[ \frac{c_D}{g_A \Lambda_{\chi}} + \frac{2}{3} \, c_3 \,
\frac{{\bf p}^2}{4m^2_{\pi}+{\bf p}^2} \nonumber \\
&+ I(\rho,P) \biggl( \frac{1}{3} \, (2c_4-c_3) + \frac{1}{6 m} \biggr)
\biggr] \,,
\label{1beff}
\end{align}
where $\rho = 2 \kf^3/(3\pi^2)$ is the density of the reference state,
$\kf$ the corresponding Fermi momentum, and $I(\rho,P)$ is due to the
summation in the exchange term,
\begin{multline}
I(\rho,P) 
=1 - \frac{3 m^2_\pi}{2 \kf^2}
+ \frac{3 m^3_\pi}{2 \kf^3} \, \mathrm{arccot} 
\biggl[\frac{m_\pi^2+\frac{P^2}{4}-\kf^2}{2 m_\pi \kf}\biggr] \\[1mm]
+ \frac{3 m^2_\pi}{4 \kf^3 P} \Bigl(\kf^2 + m_\pi^2 -\frac{P^2}{4}
\Bigr) \log \biggl[ \frac{m_\pi^2+(\kf-\frac{P}{2})^2}{m_\pi^2
+(\kf+\frac{P}{2})^2} \biggr] .
\end{multline}
The contributions from 2b currents depend on the density of the
reference state. We consider a typical range for nuclei $\rho=
0.10...0.12 \fmiq$, which corresponds to average nucleon momenta $Q
\sim 150-200 \mev$, so that 2b currents are expected to be more
important compared to very light nuclei. For these densities, we have
$I(\rho,P) = 0.64...0.66$, using the Fermi-gas mean-value $P^2 = 6
\kf^2/5$, and $I(\rho,P) = 0.58...0.60$ for $P=0$, so the 
total-momentum dependence is very weak.

The effective 1b current ${\bf J}^{\rm eff}_{i,{\rm 2b}}$ only
contributes to the GT operator and can be included as a correction to
the $g_A(p^2)$ part of the 1b current, Eq.~(\ref{Js}). This result is
general for a spin/isospin symmetric reference state. Beyond the Fermi-gas
evaluation, only the momentum dependence will be replaced by a 
state/orbital dependence.

\begin{table}
\begin{center}
\begin{tabular*}{0.48\textwidth}{l|c|c|c|cc}
\hline\hline
& & & $c_D=0$ & $q = 0.74$ & $q = 0.96$ \\
& $c_3$ & $2c_4$-$c_3$ & $q$ & \multicolumn{2}{c}{$c_D$} \\ \hline
EM & -3.2 & 14.0 & 0.72...0.66 & -0.17...-0.70 & 
-2.34...-2.51 \\
EM+$\delta c_i$ & -2.2 & 11.0 & 0.78...0.73 & 0.40$^*$...-0.11 
\hspace*{-1.5mm} & \hspace*{-1mm} -1.78$^*$...-1.92 \\
EGM & -3.4 & 10.2 & 0.80...0.75 & 0.55...0.04 & 
-1.63...-1.77 \\
EGM+$\delta c_i$ & -2.4 & 7.2 & 0.85...0.82 & 1.11...0.63 & 
-1.06...-1.18 \\
PWA & -4.78 & 12.7 & 0.75...0.69 & 0.08...-0.44$^*$ 
\hspace*{-1.5mm} & \hspace*{-1mm} -2.10...-2.26$^*$ \\
PWA+$\delta c_i$ & -3.78 & 9.7 & 0.81...0.76 & 0.64...0.14 & 
-1.53...-1.67 \\
\hline\hline
\end{tabular*}
\end{center}
\vspace*{-5mm}
\caption{Quenching $q$ of the $p=0$ GT operator
predicted by the long-range parts of 2b currents ($c_D=0$) for given
$c_3, c_4$ couplings. Dots correspond to a density range 
$\rho = 0.10...0.12 \fmiq$, and all $c_3, c_4$ values are
in GeV$^{-1}$. In addition, we give the short-range
coupling $c_D$ obtained by requiring that 2b currents lead to a 
quenching of $q=0.74$ or $q=0.96$ over this density range. For
given $q$, the $^*$ values correspond to the weakest/strongest
$p$ dependence in Fig.~\ref{GT_1b2b} (smallest/largest $|c_3| \rho$),
which yield the thin bars in Fig.~\ref{nme}.\label{table}}
\vspace*{-4mm}
\end{table}

For $c_D=0$, the leading 2b currents are fully determined by the
couplings $c_3, c_4$ from nuclear forces. We take $c_3, c_4$ from the
N$^3$LO NN potentials of Ref.~\cite{EM} (EM)
or 
Ref.~\cite{EGM} (EGM), 
as well as from a NN partial wave analysis (PWA)
extraction~\cite{PWA}. Because the $c_3, c_4$ are large in chiral EFT
without explicit Deltas, we also explore typical changes expected at
higher order $\delta c_3=-\delta c_4 \approx 1 \gevi$~\cite{RMP}. The
different $c_3, c_4$ are listed in Table~\ref{table}. As shown, the
corresponding long-range parts of chiral 2b currents predict a
quenching of the GT operator at $p=0$ ranging from $q=0.85...0.66$ for
densities $\rho=0.10...0.12 \fmiq$. The largest quenching is obtained
from the largest $2 c_4$--$c_3$ coupling, which is closest to the
single-Delta excitation with $2 c_4$--$c_3 \approx 15 \gevi$. The
density dependence of the 2b-current contributions for $c_D=0$
is also shown as inset in Fig.~\ref{GT_1b2b}.

This demonstrates that chiral 2b currents naturally contribute to the
quenching of GT transitions. A reduction of $g_A$ in the currents
is also expected considering chiral 3N forces as
density-dependent two-body interactions~\cite{Holt}.
We can constrain the short-range coupling
$c_D$ by requiring to reproduce an empirical quenching $q=0.74$ needed
in many-body calculations~\cite{WildenthalMP}. This leads to values
$|c_D| < 1$ in Table~\ref{table}, which is compatible with independent
determinations from 3N forces, especially in the best studied EM
case~\cite{Gazit,3Nfits} (the comparison should be made for large 3N
cutoffs); e.g., $c_D$'s obtained from the $^3$H half-life
fit~\cite{Gazit} lead to $q=0.75...0.67$. Because part of the
quenching may be due to truncations in the many-body basis, we can ask
what values of $c_D$ are needed to reproduce a smaller quenching
$q=0.96$ (based on an extraction $q^2=0.92 \pm 0.11$ from GT strength
functions to high energies~\cite{Sasano}). As shown in
Table~\ref{table}, the resulting $c_D$ are more negative, which
seems compatible with 3N force fits only for the EGM and PWA
cases~\cite{3Nfits}. Because of these uncertainties, we consider a
conservative range from empirical to small quenching. The resulting
quenching of $g_A$ is shown in Fig.~\ref{GT_1b2b} versus
density. Chiral 2b currents will universally affect $p \approx 0$ GT
lifetimes and strength functions by $q^2$ and \tnbb\ lifetimes by
$q^4$.

The momentum-transfer dependence is predicted in chiral EFT and shown
in Fig.~\ref{GT_1b2b} for the 2b-current contributions. This $p$
dependence is determined by the density and the $c_3$ coupling in
Eq.~(\ref{1beff}), and is stronger for larger $|c_3| \rho$
values. With increasing $p \sim m_\pi$, the impact of 2b currents, and
thus the quenching, is reduced and the GT 1b operator could even be
somewhat enhanced in the smaller quenching case. A reduction is
consistent with muon capture studies, which probe $p \sim$ $m_\mu \sim
100 \mev$, where GT quenching is not needed~\cite{Zinner}.

Finally, we apply chiral EFT currents to the \znbb\ decay NME, which
is given by the nuclear currents~\cite{Tomoda}:
\be
M^{0\nu\beta\beta} = \Bigl\langle 0^{+}_{f} \Bigl| \frac{R}{g_{A}^{2}}
\sum_{i,j} \tau_{i}^{-}\tau_{j}^{-}
\int \frac{d{\bf p}}{2\pi^{2}} \, e^{i{\bf p}\cdot{\bf r}_{ij}
} \frac{J_{i}^{\mu} J_{j,\mu}}{p (p+ \overline{E}_{\rm I})} 
\Bigr| 0^{+}_{i} \Bigr\rangle .
\ee
Here, 
a size scale $R = 1.2 A^{1/3} \,
{\rm fm}$ is introduced so that $M^{0\nu\beta\beta}$ is
dimensionless, and the closure approximation was used with an average
intermediate-state energy $\overline{E}_{\rm I}$.

Constructing the \znbb\ decay operator in chiral EFT, the nuclear
currents are given by the 1b currents $J^{0}_{i,{\rm 1b}}$ and ${\bf
J}_{i,{\rm 1b}}$ of Eqs.~(\ref{Jt}) and (\ref{Js}), and we include
2b currents at the normal-ordered 1b level ${\bf J}^{\rm
eff}_{i,{\rm 2b}}$ of Eq.~(\ref{1beff}). With these, the transition
operator can be decomposed into GT, Fermi (F), and tensor (T) terms that
receive contributions from the $V, A, P, M$ components in the
currents~\cite{Simkovic}:
\begin{align}
&J_{1}^{\mu} J_{2\mu} = h^{\rm GT}(p^2) \, {\bm \sigma}_1 \cdot {\bm \sigma}_2 
- h^{\rm F}_{VV}(p^2) - h^{\rm T} \, {\bf S}_{12}(\widehat{\bf p}) \,, \\[1mm]
&h^{\rm GT}(p^2) = h_{AA}^{\rm GT} + h_{AP}^{\rm GT} + h_{PP}^{\rm GT} +
h_{MM}^{\rm GT} \,.
\end{align}
At leading order $Q^0$, only the $p=0$ axial-axial and vector-vector
terms $h_{AA}^{\rm GT}(0)$ and $h_{VV}^{\rm F}(0)$ contribute. At
order $Q^2$, this includes the $p^2$ terms in $g_{V,A}(p^{2})$ and the
other terms in the 1b currents. The $Q^2$ terms are also important in
phenomenological currents, as they lead to short-range correlation
effects being small~\cite{short}, which is consistent with chiral
EFT. At order $Q^3$, 2b-current contributions enter in
$h_{AA}^{\rm GT}$ and $h_{AP}^{\rm GT}$. The $h^{\rm T}$ parts are
small (less than $1 \%$, except in
$^{48}$Ca)~\cite{Menendez,Kortelainen}, consistent with neglecting
$({\bf k} \cdot {\bm \sigma} \, {\bf k} - \frac{1}{3} k^2 {\bm
\sigma})$ terms from the 2b currents.

We then calculate $M^{0\nu\beta\beta}$ based on chiral EFT currents
for $^{48}$Ca, $^{76}$Ge, $^{82}$Se, $^{124}$Sn, $^{130}$Te and
$^{136}$Xe. These are challenging calculations and require
approximations. Here we perform state-of-the-art SM calculations using
the coupled code NATHAN~\cite{Caurier}, ideal for $0^{+}$ states. The
details of the SM spaces and interactions are given in
Ref.~\cite{Menendez}. Because 2b currents are independent of the
many-body spaces, they should be included in all calculations, and we
focus on relative changes.

Our results are shown in Fig.~\ref{nme} for successive orders in
chiral EFT. At order $Q^2$, the NMEs are similar to calculations based
on phenomenological 1b currents only, with differences mainly due to
the $Q^2$ terms compared to phenomenological dipole form factors, as
well as different $\Lambda_A$ values used. At order $Q^3$, the
contributions from chiral 2b currents range from a $10\%$ increase to
a $35\%$ reduction of $M^{0\nu\beta\beta}$. This range is obtained for
the couplings of Table~\ref{table} and overlaps with the 
prediction based on the long-range parts of 2b currents. The
changes at order $Q^3$ agree with the $p \sim 100 \mev$ estimate from
Fig.~\ref{GT_1b2b}. The lower $M^{0\nu\beta\beta}$ values correspond
to the quenching of $g_A$ in SM calculations of $p \approx 0$
single-$\beta$ and \tnbb\ decays, but the $p$ dependence weakens the
quenching compared to the naive GT $q^2 \approx 0.74^2 \approx 0.55$
or $45\%$ reduction. Finally, we note that in Delta-full EFT, the
long-range single-Delta part of 2b currents would contribute at order
$Q^2$, which would shift the $Q^2$ results more into the $Q^3$ range,
improving the order-by-order expansion in Fig.~\ref{nme}.

This presents the first study of chiral 2b currents for GT transitions
and \znbb\ decay in medium-mass nuclei. Compared to light nuclei,
their contributions are amplified because of the larger nucleon
momenta. For a spin and isospin symmetric reference state, the leading
axial 2b currents contribute only to the GT
operator. A quenching of $p \approx 0$ GT transitions is predicted by
the long-range parts of 2b currents, and we have estimated the
theoretical uncertainty from the short-range part assuming an
empirical to small quenching range.

Chiral EFT predicts the momentum-transfer dependence of 2b currents
that is probed in \znbb\ decay. This first calculation of the \znbb\
decay operator based on chiral EFT currents shows that 2b
contributions to the NMEs range from $-35 \%$ to $10 \%$ and should be
included in all calculations. Future directions are: going beyond the
1b approximation, investigating other electroweak transitions,
developments of consistent interactions and operators (see also~\cite{Holt}), 
and astrophysics explorations.

\begin{acknowledgments}
We thank R.\ J.\ Furnstahl, T.\ R.\ Rodr{\'i}guez and U.\ van Kolck for
very useful discussions. This work was supported in part by the DFG
through grant SFB 634 and the Helmholtz Alliance HA216/EMMI.
\end{acknowledgments}


\begin{thebibliography}{99}
\bibitem{dipole} The same structure is obtained
phenomenologically based on symmetries~\cite{Tomoda}, with dipole-form
couplings.
\bibitem{vectortemp} Vector temporal parts do not contribute
at this order~\cite{Park}.
\end{thebibliography}
\end{document}